# Experimental investigation on the performance of thermosyphon charging of a single-medium stratified storage system for concentrated solar power applications

Dipti Ranjan Parida, Saptarshi Basu[*], Dhanush A P

Department of Mechanical Engineering, Indian Institute of Science, Bangalore, India-560012

## Abstract

Concentrated solar power (CSP) plants utilize two-tank, sensible-heat thermal energy storage (TES) for uninterrupted electricity generation. However, the cost for the design and operation of TES is expensive. Therefore, researchers are focusing on implementing single-tank storage. Additional cutbacks can be made by utilizing pump-less thermosyphon charging for the TES. But prior thermosyphon researches for TES are related to domestic water-heating systems of small-capacity (<100 liters) and low-temperature (<100 °C). Thus, investigations into thermosyphon charging for high-temperature storage are desired.

This study focuses on thermosyphon-charging and storing of a single-medium stratified TES. The experiments were conducted on a 370 liters cylindrical storage (aspect ratio 4:1) with a heat-pipe system (3-liter volume) acting as a collector. Dowtherm-A oil was used as the heat transfer fluid (HTF), and the thermal expansion of HTF was accommodated in an expansion tank via two different designs (top and bottom connections from storage tank to expansion tank). Moreover, continuous and pulsatile charging are investigated for low (150 °C) and high (250 and 300 °C) temperatures. The results indicate that the maximum HTF temperature coming out of the heating pipes is ~25 °C more for the bottom-expansion design. Furthermore, it results in higher charging efficiency than the top-expansion setup for high-temperature studies. Finally, it is revealed that under design conditions, there are limits on the degree of thermal stratification achieved in the charging cycle and the maximum layover time allowable for interrupted charging. These results provide insights into the operational strategy of thermosyphon-charging stratified storage for CSP applications.

**Highlights**

- Thermosyphon charging for high-temperature stratified TES is examined for concentrated solar power applications.
- Results show that accommodating HTF externally through a bottom-expansion connection is beneficial for thermosyphon charging.
- There is a design limit to the degree of thermal stratification resulting from thermosyphon charging.
- The degradation of thermal stratification and exergy analysis determines the maximum layover period for successive charging.

---

[*] Corresponding author.
  Email address: sbasu@iisc.ac.in (S. Basu)

**Word count**: 5600

**Nomenclature**

**Abbreviation**

CSP: Concentrated solar power

TES: Thermal energy storage

STSHS: Single tank sensible heat storage

SMSHS: Single medium sensible heat storage

DMSHS: Dual medium sensible heat storage

HTF: Heat transfer fluid

**Symbols**

$D$: Diameter of the storage tank, m

$H$: Height of the storage tank, m

$z$: Vertical height of thermocouples, m

$Z^*$: Normalized height

$\rho$: Density of HTF, $\left(\frac{kg}{m^3}\right)$

$\rho_c$: Density of cold HTF

$\rho_h$: Density of hot HTF

$\rho_i$: Density of i$^{th}$ zone

$T$: Temperature reading of thermocouple, $K$

$T_c$: Temperature of cold HTF

$T_h$: Temperature of hot HTF

$T^*$: Normalized temperature

$T_{avg}$: Average temperature of the storage

$T_{inlet}$: Inlet HTF temperature

$T_{ini}$: Average initial temperature of the storage

$T_{i,t}$: Temperature corresponding to i$^{th}$ zone

$t$: Time, $minute$

$\tau$: Normalized time ($\frac{t}{60}$), *hour*

$T_0$: Ambient temperature (25 °C)

$C_p$: Specific heat capacity of HTF, $\left(\frac{kJ}{kgK}\right)$

$\kappa$: Thermal conductivity of HTF, $\left(\frac{W}{mK}\right)$

$\eta(t)$: Charging efficiency

$A_t$: Atwood number, $A_t = \frac{\rho_c - \rho_h}{\rho_c + \rho_h}$

$A_t(JH)$: Atwood number corresponding to jacketed-heater

$A_t(IN)$: Atwood number corresponding to intet

$A_t(TS)$: Atwood number corresponding to thermal storage

$ISI$: Ideal stratification index

$n$: Number of zones

$V_i$: Volume of i$^{th}$ zone

$C_{p_i}$: Specific heat capacity of i$^{th}$ zone

$\theta_j^*$: Equivalent normalized temperature

$f(\theta_j^*)$: Ideality factor

$\xi_t$: Exergy, *kJ*

$\xi_0$: Initial exergy of TES

$\xi^*$: Normalized exergy

# 1. Introduction

Electricity is ubiquitous in the modern-day World, and a significant portion of the World's electricity is produced by burning fossil fuels. However, fossil fuels will eventually run out. Hence, alternative electricity generation technologies for renewable energy resources are encouraged. One promising technology is the Concentrated Solar Power Plant (CSP). It is a thermal power plant technology that converts solar thermal energy to electricity by concentrating solar radiation. Depending on the types of concentrating system, the solar thermal harvesting system is classified as i) power tower system, ii) parabolic trough system, iii) parabolic disc system, and iv) linear fresnel system [1]. Although solar energy is abundant, CSPs necessities additional thermal energy storage (TES) for uninterrupted electricity generation owing to the time-dependent nature of the source. The TES system is generally categorized as i) thermo-chemical heat storage, ii) latent heat storage, and iii) sensible heat storage [2]. The sensible-heat storage is more appreciated than the rest as it is low-cost, reliable, and matured technology; it can be operated at higher temperatures (~600 °C) [3]. From the design perspective, two types of sensible heat storage are present, i) single-tank storage and ii) two-tank storage. At present, two-tank sensible heat storage is already made operational in CSPs. However, single-tank storage offers a low-cost alternative (35-48% less) than two-tank storage [4,5] and has been the subject of investigation for the last decade. A detailed classification of concentrated solar power technology is depicted in Figure 1.

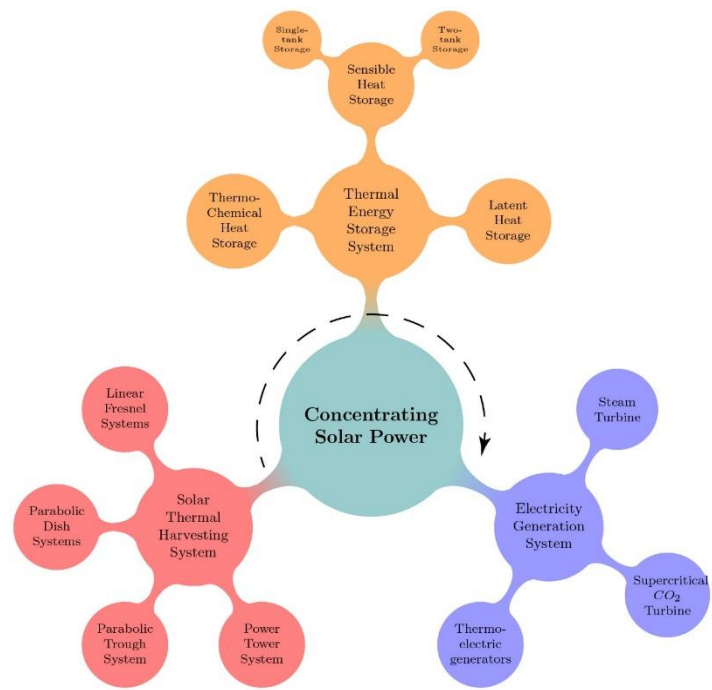

Figure 1: Classification of concentrating solar power technologies. Each CSP comprises a solar thermal harvesting system and an electricity generation system. The thermal energy storage system is generally installed in large-size plants to reduce the gap between electricity production and demand (except for the parabolic disc system).

In single-tank sensible heat storage (STSHS), thermal energy is preserved either in a liquid or in a combination of liquid and solid filler material, like rock and sand. Without the filler material, the liquid medium (usually consisting of oil/molten-salt depending on the operational temperature) acts as both heat transfer and storage medium. In that case, the STSHS is called single-medium sensible heat storage (SMSHS). Else, the solid filler materials act as the primary heat storage medium and the liquid act as the heat transfer medium. Then, the STSHS is called dual-medium sensible heat storage (DMSHS) [6]. From the operational point of view, a complete thermal cycle of STSHS consists of three periods: a) the charging period, b) the storage period, and c) the discharging period, and the STSHS operates on the principle of thermal stratification. The thermal stratification layer is known as the thermocline [7], which forms due to the thermal mixing of hot and cold HTF at the beginning of the charging and discharging periods. Depending on charging or discharging, it then translates towards the bottom or top of the TES tank. Though thermocline prevents further mixing of hot and cold HTF, its thickness can increase due to very high and/or low charging-discharging rates. At high charging/discharging, the inertia of the incoming fluid creates significant mixing, whereas, at low charging/discharging, prolonged thermal diffusion increases the thermocline thickness. Moreover, the thermocline broadens due to additional temporal thermal interactions between hot and cold HTF within the TES. These are intrafluid convection, wall conduction, and convective loss to ambient (Figure 2).

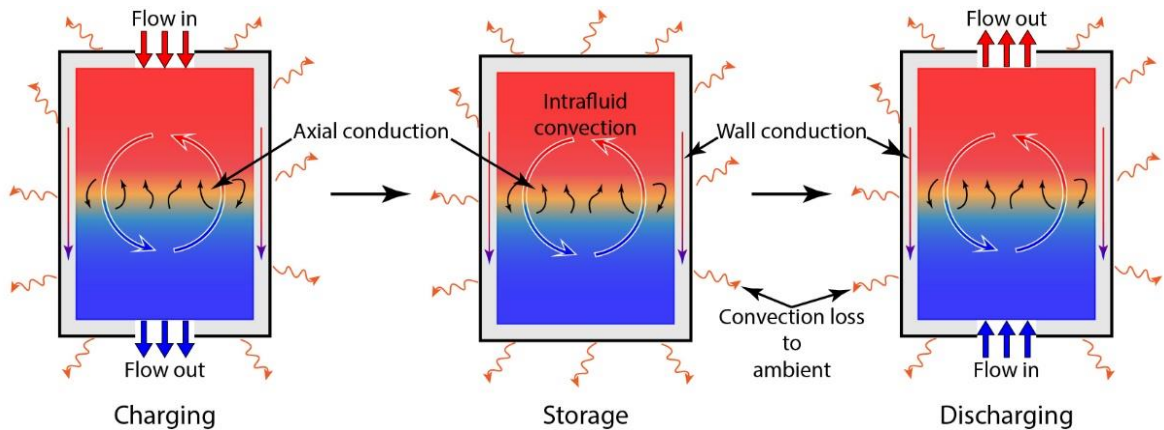

Figure 2: Thermal interactions in the single-tank sensible heat storage system during charging, storage, and discharging periods. During charging, the hot HTF (red-colored) is poured from the top, and the cold HTF (blue-colored) is taken away from the bottom of the TES tank simultaneously. During discharging, the flow is reversed. The stratified thermal zone (thermocline) is shown in the middle between hot and cold HTF.

Several numerical and experimental studies have been carried out for DMSHS [8–12]. Pacheco et al. [13] performed experiments on molten-salt thermocline TES with quartzite and silica filler materials. They reported that the cost of TES can be reduced by 1/3 compared to two-tank TES. Brosseau et al. [14] examined the filler materials for thermocline TES with HitecXL molten salt (a ternary mixture of 44 wt. % $CaNO_3$, 12 wt. % $NaNO_3$, and 44 wt. % $KNO_3$). They reported a significant $CaCO_3$ scale built up in the storage tank, which further increases if the operating temperature is more than 450°C. Thus, it indicates that the filler materials limit the maximum operating temperature of HTF in DMSHS. Moreover, scaling/fouling will alter the thermophysical properties of primary HTF (oil/molten-salt) and thus degrade heat transfer characteristics and pumpability, creating operational and maintenance challenges.

In a numerical investigation [15], it is reported that the discharge efficiency of the dual-medium thermocline storage decreases with an increase in Reynolds number, which makes it less likely that DMSHS can be suitable during high demand. Furthermore, [16] reports that asbestos filler material performs better than alumina spheres in terms of charging-discharging efficiency and thermocline thickness in oil-based TES. However, the volumetric heat capacity of asbestos is comparatively lower than that of alumina. This signifies that the TES tank size of asbestos-based storge will be significant for an equivalent heat storage capacity. To summarize, though DMSHS reduces the material cost of the liquid heat transfer medium, it creates additional challenges like an improper pairing of filler-HTF materials, greater thermal diffusion, thermal ratcheting, maintenance issues, and so on [12,16–18]. Nevertheless, the SMSHS, which is free from these challenges, is rarely studied in the literature for high-temperature applications.

Regarding the discharging-discharging process for SMSHS, a forced flow of HTF is necessary for the discharging cycle (via pump drive) as the heat extraction rate varies depending on the demand. However, the charging of SMSHS can be carried out with either forced or natural circulation of HTF. The natural circulation of HTF is gravity driven and is known as the thermosyphon effect [19]. It is a half-century old technique and has a wide range of applications, including cooling (nuclear reactors, gas turbine blades, internal combustion engines, electronics, and so on) and heat-extraction (geothermal, waste-heat, and solar water heating) [20,21]. Depending on the phase of the HTF, two types of thermosyphon effect are present: a) single-phase thermosyphon and b) two-phase thermosyphon. In the case of SMSHS, the HTF does not change its phase; hence, single-phase thermosyphon charging is of interest. As no pump is required for thermosyphon charging of the HTF, it'll reduce both the instrument and operational cost of the TES. This passive way of charging SMSHS has been studied significantly in the past and has been commercially applied to solar domestic hot water TES systems [22]. Such studies, however, give a limited perspective of thermosyphon-charging as the working fluid is water, the operating temperature is less than 100 °C, and the volume of TES is usually less than 100 liters. Moreover, it is reported in [23] that the dimensions and thermal power of a thermocline TES determines its maximum efficiency, and it is expected that the large/real-sized tanks will behave differently than that of small/prototype-sized TES. So, it is not firm how the passive charging will behave for high-temperature, large-scale TES systems for CSP plants. In this regard, the following queries can be asked for thermosyphon charging of large-scale SMSHS:

For a given volume of the storage tank and heating volume (residing volume of the heat collection pipes) of the HTF,

  i. What will the thermal-stratification profile be in the thermocline storage as the charging cycle evolves?
  ii. What will be the efficiency of thermosyphon charging?
  iii. Considering the large volume of HTF and high operating temperature, the thermal expansion of HTF will be significant. So, how do different ways of accommodating thermal expansion affect the thermal stratification inside the TES?
  iv. What will be the operation philosophy for thermosyphon charging of high-temperature storage?

In a nutshell, the efficacy of the CSPs directly depends upon both the cost and efficiency of TES, and hence investigations for an efficient thermosyphon-linked high-temperature STSHS need to be carried out.

This brings us to the objective of this paper, which is to understand how to utilize thermosyphon charging in a single-medium, single-tank, large-scale sensible heat storage for concentrated solar power applications. In anticipation of this, we developed a thermocline storage system having a storage capacity of 370 liters, which was linked to a jacketed heater for experimental investigation. The thermal expansion of the HTF was accommodated in an expansion tank via two different connections, and the thermosyphon charging was carried out for continuous and pulsating charging with the help of a control valve. In total, eight experiments were conducted, each having a charging time of 12 hours followed by storage of 24 hours, to examine the evolution and degradation of thermal stratification inside the TES.

## 2. Experiments
### 2.1. Experimental setup

Figure 3 shows the schematic of an experimental rig developed under the IMPRINT India initiative at IISc, Bangalore. It is a hybrid TES system that integrates sensible and latent heat storage followed by a water-based heat extraction system. It comprises two HTF circuits (HTF-1 and HTF-2) and six independent thermal cycles/process modes, including both charging and discharging (see Appendix, section). The thermosyphon test loop, investigated in this work, is the HTF-1 part of this experimental rig, which contains (i) a sensible heat storage tank (thermocline tank), (ii) an expansion tank, (iii) a jacketed heater, and associated pipings. The thermocline tank (see Figure 4) is cylindrical with an aspect ratio $\frac{H}{D} \approx 4$, where $H = 1.9\ m$ is the height and $D = 0.498\ m$ is the diameter of the storage tank; the storage volume of TES is 370 liters. It comprises several manifolds for HTF transport with respect to the jacketed heater and the expansion tank. The HTF transport manifolds are branched out to 6 hemispherical diffuser arrangements (both at the top and the bottom of the TES tank) to reduce the thermal blending of hot and cold HTFs. Moreover, polymer bush layers are provided between thermocline storage and the support structures to minimize additional heat loss (see Figure 4).

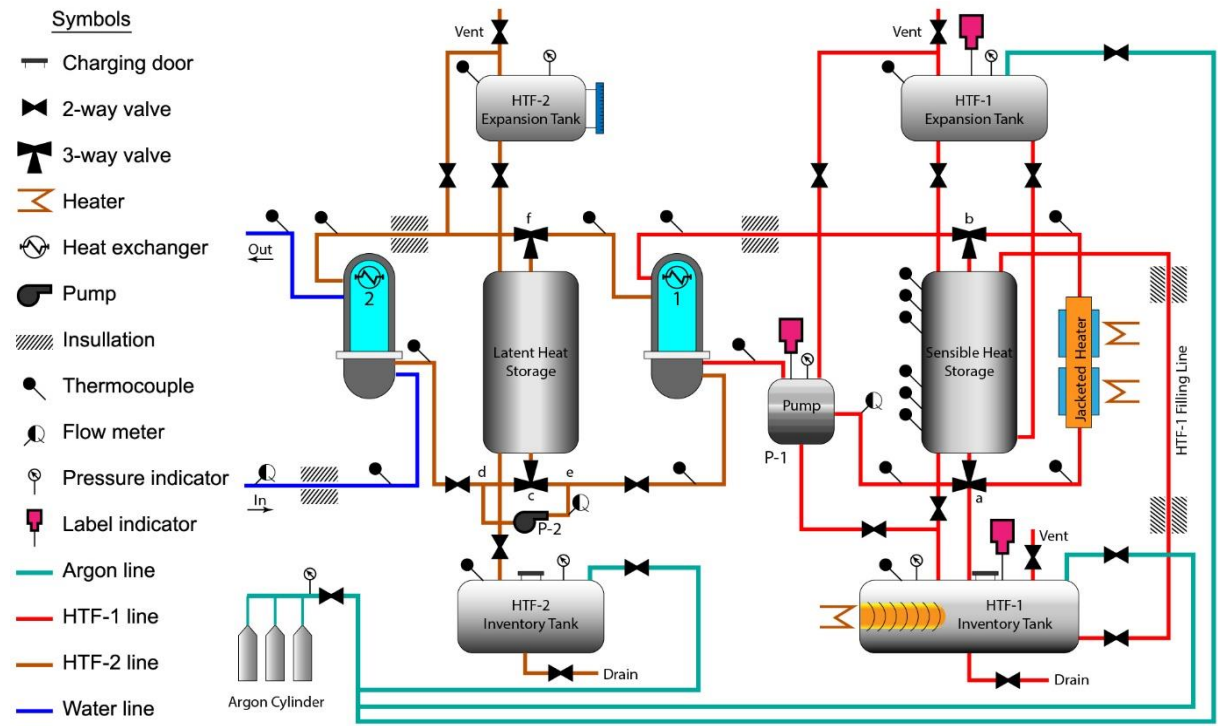

Figure 3: Schematic of the hybrid TES developed under the Imprint India initiative. It consists of high-temperature sensible heat storage without filler material, medium-temperature latent heat storage, and a water-based heat extraction system. The HTF-1 can be oil/molten salt, and the SHS tank is charged via thermosyphon.

The jacketed heater is coupled with the thermocline storage through a charging HTF pipeline to replicate the concentrated solar energy input. The charging HTF line is divided into three heat pipes inside the jacketed heater. The length and diameter of these heat pipes are 2 m and 0.0254 m, respectively. So, the ratio of the heating volume to storage volume for thermosyphon charging is $\frac{3.04\ lt.}{370\ lt.} = 0.008$. These heat pipes are inclined 30° to the vertical in accordance with the literature [24]. The remaining pipes between thermocline storage, jacketed heater, and expansion tank are of the same bore diameter (0.0254 m). Moreover, the complete loop is wrapped by 50 mm glass wool insulation with the help of aluminum cladding/foil.

A series of thermocouples (24 numbers) are installed vertically in the TES tank (with a gap of 76 mm between each) to measure the thermal stratification. Also, two thermocouples are installed at the bottom and top of the jacketed heater to measure the temperature of incoming and outgoing HTF. All these thermocouples are of K-type and have an uncertainty of Max($\pm 2.5$ °C or 0.75%). The temperature readings are taken to a PLC monitor though thermocouples-modules provided by the manufacturer (HEATCON Pvt. Ltd., Bangalore, India). Furthermore, the pneumatic valves and heaters are operated through PLC. The experimental loop is shown in Figure 5.

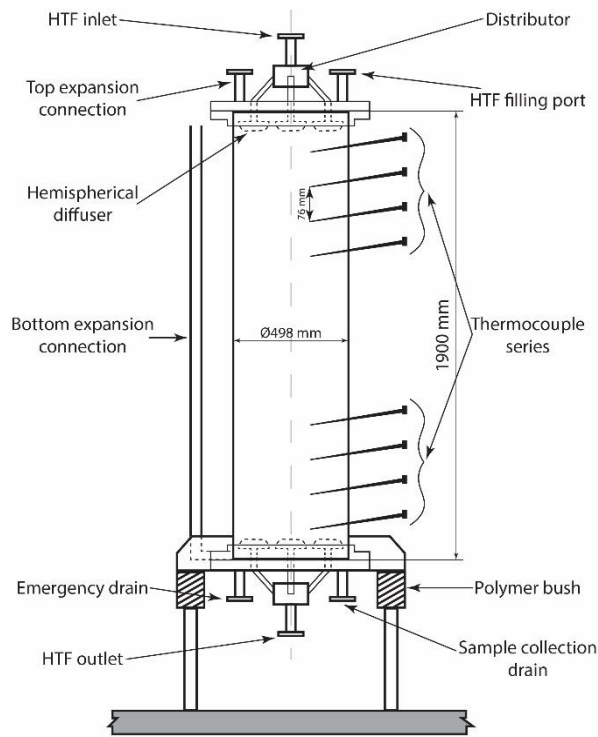

Figure 4: Schematic of the thermocline storage tank showing thermocouple array. Polymer bush is provided at the bottom of the storage to minimize heat loss. The top and bottom distributors consist of six hemispherical diffusers (see Appendix).

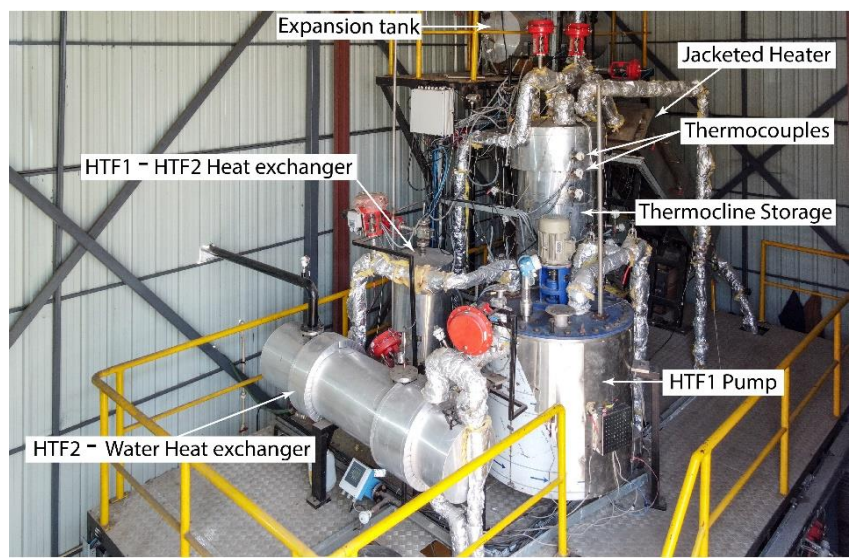

Figure 5: Experimental TES setup built at IISc. Dowtherm-A oil is used as the HTF-1 for the thermosyphon charging experiment; (The PCM storage is not shown in the figure).

## 2.2. HTF materials and properties

Dowtherm-A oil is used as the heat transfer fluid for this investigation as it is cheaper than Syltherm 800, which is generally used in parabolic trough systems. Moreover, its operating temperature range of 15-400 °C is comparable to that of Syltherm 800 (-40 to 400 °C). More importantly, it has a lower fluid expansion of ~56% than Syltherm 800 (~72%) in the range of 25-400 °C [25]. The Dowtherm-A HTF used in this experiment is procured from Thermic Fluids Pvt Ltd., Mumbai, India. The thermophysical properties given by the manufacturer are curve fitted (see Appendix), and the polynomial expressions are given below.

$$\rho = 1112.9 + 0.8779T - 0.0058T^2 + 9 \times 10^{-6}T^3 - 6 \times 10^{-9}T^4 \quad \left(\frac{kg}{m^3}\right) \tag{1}$$

$$C_P = 0.7002 + 0.0029T \quad \left(\frac{kJ}{kgK}\right) \tag{2}$$

$$\kappa = 0.1856 - 0.0002T \quad \left(\frac{W}{mK}\right) \tag{3}$$

where $\rho \left(\frac{kg}{m^3}\right)$, $C_P \left(\frac{kJ}{kgK}\right)$, $\kappa \left(\frac{W}{mK}\right)$ are the density, the specific heat capacity, and the thermal conductivity of the saturated Dowtherm-A liquid in the range of 15-400 °C. $T$ is the temperature in $K$.

## 2.3. Operational procedures and experimental conditions

One of the main operational challenges in large-scale, high-temperature sensible heat storage is how to accommodate significant HTF expansion. The top portion of the storage tank is generally left unfilled to accommodate the expanded HTF. Though this method is suitable for solar water heating systems, it may not be appropriate for high-temperature storage as

i. The HTF (oil/molten-salt) needs to be operated in an inert atmosphere, like Nitrogen or Argon, to avoid air contact that accelerates thermal decomposition; thus, an inert-gas inlet has to be provided at the top of the storage.
ii. A pressure gauge and relief valve need to be installed to monitor and control the surge in pressure at the top of the storage tank.
iii. A level indicator is also needed to monitor the HTF level in the storage and so on.

These additional modifications/instrument installation will create design challenges and make the top of the TES tank clumsy and inaccessible. This challenge can be evaded by external accommodation for the HTF in an expansion tank. However, the expansion tank can be joined to either the storage tank's top or the bottom. Both of these expansion tank connections are examined in this study.

For the experiment, the thermosyphon charging loop (which contains the thermocline tank, jacketed heater, and expansion tank) is purged by argon and filled with HTF (Dowtherm A). The jacketed heater is then set to a fixed temperature (150, 250, or 300 °C). When the jacked heater's skin temperature reaches the set temperature, its power is decreased to 70%; when the skin temperature drops by 10 °C, its power is raised to 100%. This automated PLC-controlled jacketed heater resulted in a ±10 °C variation in the skin temperature. Two different ways of charging performed were i) continuous type and ii) pulsating type. In continuous charging, the valve-b (see Figure 3) is left open for thermosyphon charging. In pulsating charging, the valve-b is opened and closed periodically (15 minutes open followed by 15 minutes closed). The experimental study cases include low (150 °C) and high (300 °C) temperature charging for continuous-type charging. However, it is revealed from trial runs that the pulsating charging at low temperature is not beneficial as a relatively lower gain in temperature during the pulsating diffused owing to slower thermosyphon charging. Thus, the pulsating-type charging is tested for 250 and 300 °C. The detailed study cases are given in Table 1.

Table 1: Study cases for thermosyphon charging

| Test cases | Expansion tank connection | Charging Type | Set Temperature (°C) | Set Atwood Number ($A_t$) |
|---|---|---|---|---|
| Case #1 | Top | Continuous | 150 | 0.059 |
| Case #2 | Top | Continuous | 300 | 0.144 |
| Case #3 | Top | Pulsating | 250 | 0.110 |
| Case #4 | Top | Pulsating | 300 | 0.145 |
| Case #5 | Bottom | Continuous | 150 | 0.056 |
| Case #6 | Bottom | Continuous | 300 | 0.141 |
| Case #7 | Bottom | Pulsating | 250 | 0.114 |
| Case #8 | Bottom | Pulsating | 300 | 0.145 |

## 3. Results and discussion

The results obtained in this experiment are categorized into two parts, i) the charging period and ii) the storing period. First, the inlet-outlet conditions of the jacketed heater, the development of stratified profiles, and charging efficiency are discussed for the charging. Subsequently, the temporal degradation of thermal stratification and an exergy-based comparison are reported for the storage period. Also, it is essential to note that each temperature measurement is associated with some uncertainty (as mentioned in section 2.1 ); however, the error bars are not included in the temperature profiles for clarity.

### 3.1. Charging period
#### 3.1.1. Inlet-outlet conditions

Figure 6 shows the temperature profiles of the HTF at the outlet of the jacketed heater for all study cases. For the continuous charging (Figure 6; A and C), a maximum temperature of 98.6 °C (case #1) and 180.3 °C (case #2) is achieved for top-expansion connection, whereas 101.7 °C (case #5) and 206.4 °C (case #6) is obtained for bottom-expansion connection. So, a temperature difference of ~3 °C and ~26 °C is present between the top and the bottom expansion connection arrangements for the low (150 °C; cases #1 and #5) and high (300 °C; cases #2 and #6) temperature charging cases.

For pulsating charging, interim peaks in temperature profiles are shown in Figure 6, B and D. However, these peaks are not regular and possibly associated with/caused by the actual temperature control of the jacketed heater. Regardless, the maximum temperatures obtained in these study cases are 180.9, 213.7, 183.9, and 242.5 °C for cases #3, 4, 7, and 8, respectively. Excluding the peak temperatures, the top-expansion setup achieved a maximum temperature of nearly 162 °C and 180 °C for cases #3 and 4, respectively. In comparison, the bottom-expansion setup reached a maximum temperature of around 174 (case #7) and 205 °C (case #8). Thus, the temperature difference of ~12 °C and ~25 °C is present between these expansion setups for the pulsating charging. Assuming isothermal condition is maintained reasonably well in the jacketed heater, the following arguments can be made based on the discussed results:

   i.  The temperature gain during the thermosyphon-charging depends on the type of expansion tank connection.
   ii. The pulsating type thermosyphon-charging induces a significantly higher temperature rise in the HTF (including the interim temperature peaks) and can be valuable for high-temperature storage applications.

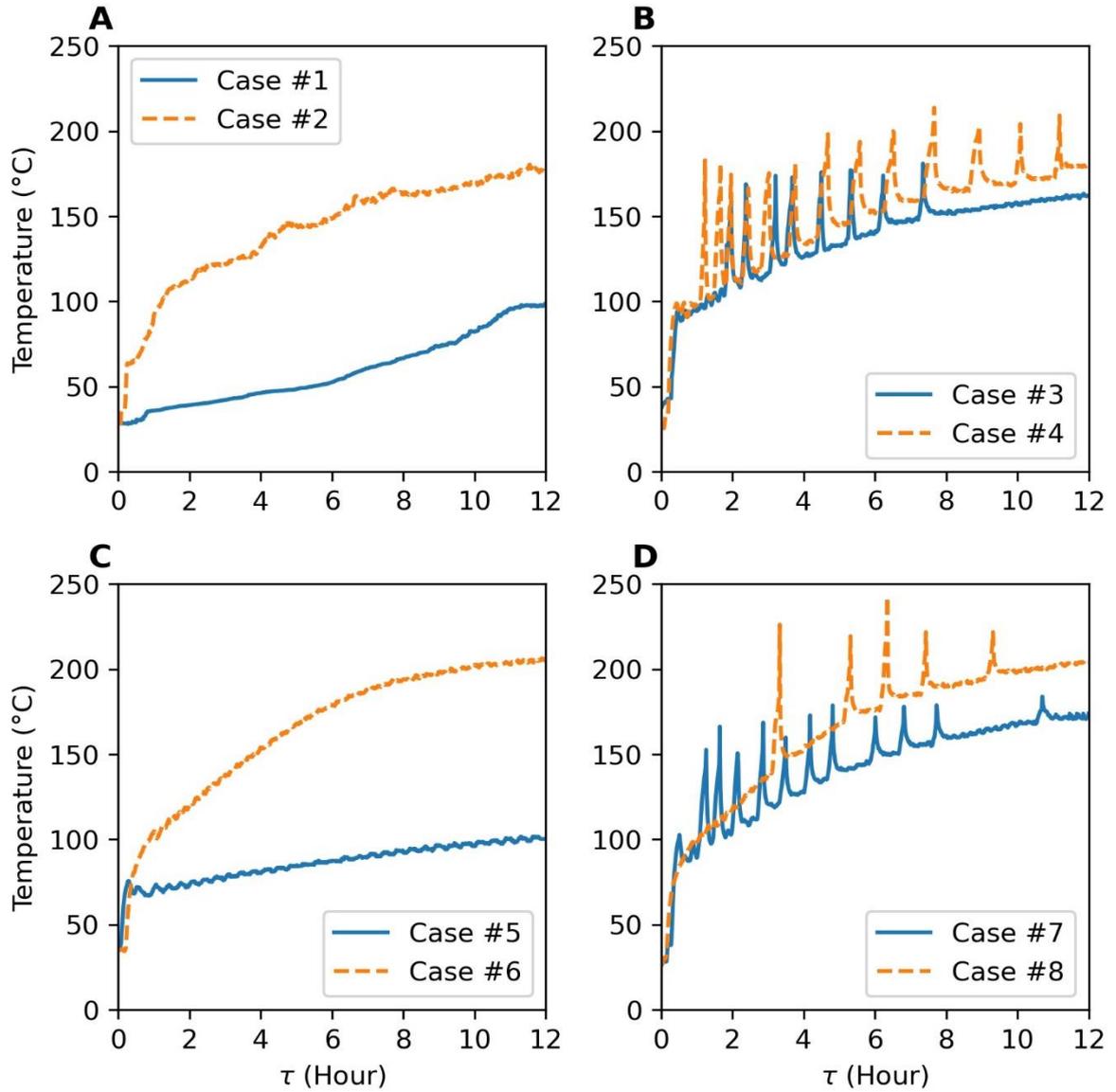

Figure 6: Temperature profiles of the HTF at the outlet of the jacketed heater; (A) and (B) are the case studies for the continuous and pulsating charging of the top expansion tank connection; (C) and (D) are the case studies for the continuous and the pulsating charging of bottom expansion tank connections, respectively.

Figure 7 shows the thermal conditions of the HTF at the jacketed heater inlet. The initial HTF temperature for the study cases varied between 25-40°C despite having a considerable cooling time of 3 days between successive experiments because the volume of the thermocline storage is reasonably large. Regarding the temporal evolution of HTF, nearly flat temperature profiles are observed for the low temperature (150 °C) charging (cases #1 and #5). However, for the high temperature (300 °C) studies, the temperatures of the HTF increased linearly after 5 hr of charging (cases #2, #4, #6, and #8) irrespective of top or bottom expansion as well as continuous or pulsating charging. The potential cause of this trend could be attributed to the higher temperature difference ($\Delta T$) between the hot and cold HTF in the storage tank. A higher $\Delta T$ will induce larger density difference, which in turn, will improve the buoyancy-driven flow rate inside the thermal loop. Hence the thermocline will travel fast towards the bottom of the storage. Moreover, the thermal diffusion in the storage tank and wall conduction will be more significant at higher $\Delta T$. In result, the cold fluid temperature rises quickly. For the 250 °C pulsating charging studies (cases #3 and #7), the temperature profiles lie between low and high-temperature study cases.

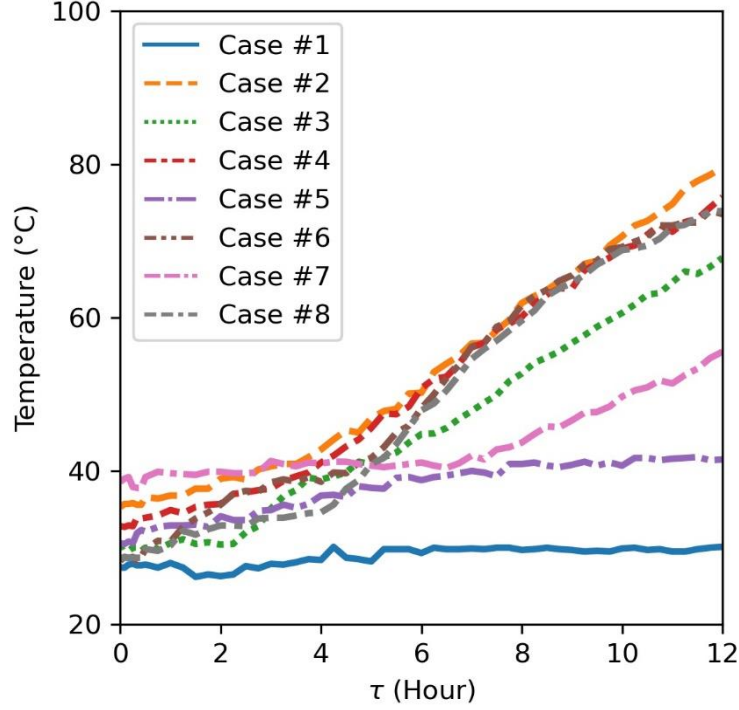

Figure 7: HTF temperature profiles at the inlet of the jacketed heater for all experimental study cases (#1-8).

### 3.1.2. Atwood number depletion

One important aspect to discuss for the thermosyphon charging is the Atwood number ($A_t$) of the thermal stratification. It is a non-dimensional number represented by the densities of hot ($\rho_h$) and cold ($\rho_c$) HTF ($A_t = \frac{\rho_c - \rho_h}{\rho_c + \rho_h}$), and an indirect measure of buoyancy. Since the density of HTF is temperature dependent, the higher the $A_t$, the higher the effective thermosyphon charging. In this study, the $A_t$ of the HTF is calculated in three categories: $A_t(JH)$, $A_t(IN)$, and $A_t(TS)$ with respect to the jacketed heater, the inlet of the storage tank, and the thermocline storage, respectively. The $\rho_h$ for these $A_t$ number categories are considered from the set skin temperature of the jacketed heater and the maximum temperature of the HTF at the inlet and the thermocline tank. Whereas the $\rho_c$ values are considered from the minimum temperature of the HTF at the beginning of the thermal cycles. Note that the $A_t(JH)$ is merely the set Atwood number for experiments as specified in Table 1. Consequently, the difference in these $A_t$ numbers signify the thermal loss associated with each step. A comparison in $A_t$ variation for the study cases is shown in a bar graph in Figure 8. Three important traits evidenced from $\{A_t(JH) - A_t(IN)\}$ are:

i. The drop in $A_t$ increases with an increase in skin-temperature of the jacketed heater (an average of 0.022 for Case #1 and 5; 0.037 for Case #3 and 7; and 0.052 for Case #2, 4, 6, and 8).
ii. The drop in $A_t$ for the top-expansion setups (cases #1, 2, 3, and 4) are relatively larger compared to the bottom-expansion setups (cases #5, 6, 7, and 8).
iii. The pulsating type thermosyphon charging is relatively more favorable than that of continuous charging as it minimizes the initial $A_t$ loss.

Regarding $A_t(TS)$, it is always lower than $A_t(IN)$ due to thermal blending of hot and cold HTF in the thermocline storage and associated heat loss. However, the difference $\{A_t(IN) - A_t(TS)\}$ increases significantly for the high-temperature studies (~0.037) compared to low-temperature studies (~0.003) as higher $\Delta T$ accelerates both thermal diffusion and wall conduction. Additionally, Figure 8 shows that the resulting $A_t(TS)$ in all cases are about $\pm 0.045$ regardless of the type of expansion connection, charging, and temperature settings. It signifies that the final thermal stratification in the SMSHS is constrained by the design of the thermosyphon loop, and it is believed to be associated with the volume ratio ($\frac{Heating\ volume}{Storage\ volume}$).

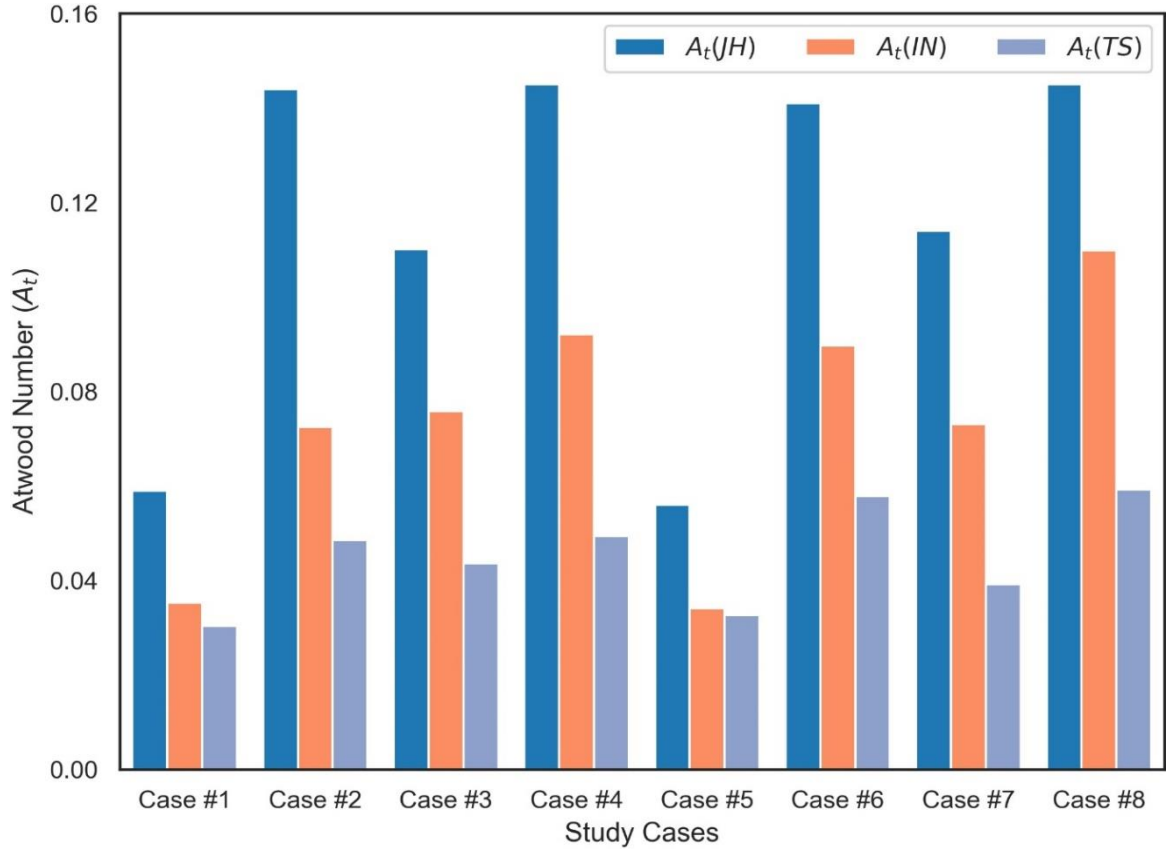

Figure 8: Atwood number variations for the study cases. $A_t(JH)$, $A_t(IN)$, and $A_t(TS)$ are Atwood numbers calculated corresponding to the jacketed heater, the inlet of the storage tank, and the thermocline storage, respectively. The drop in the Atwood number signifies the loss in stratification strength.

### 3.1.3. Thermal stratification

Figure 9 shows the evolution of stratified temperature profiles inside the thermocline storage for intermediate time instances; $\tau = 4, 6,$ and 8 horus. The corresponding temperature readings and vertical locations of the thermocouples are normalized by

$$T^* = \frac{T - T_c}{T_h - T_c} \quad (4)$$

$$Z^* = \frac{z}{H} \quad (5)$$

where $T$, $T_c$, and $T_h$ represent thermocouple temperature, the cold HTF temperature, and the hot HTF temperature, respectively; $z$, $H$ are the height of thermocouples and the thermocline storage, respectively.

The very first observation from Figure 9 is that the temperature profiles corresponding to low-temperature charging (cases #1 and 5) are nearly sigmoid, whereas the profiles are relatively linear for the rest of the study cases. It signifies that both axial thermal diffusion and wall conduction are prominent for high-temperature charging. Moreover, the temperature profiles translate from bottom to top as time progress. For instance, the topmost thermocouple temperature ($T^*$) at $\tau = 4$ is ~0.75 which then increased to ~1 at $\tau = 8$. Furthermore, a sharp linear zone near the inlet ($Z^* = 0.8$ to 1) is exist during the charging for all cases. These results suggest that the present thermosyphon loop has undergone multi-pass charging. Ideally, a sigmoid temperature profile is expected in the stratified storage, and the charging should be completed in a single pass. It is believed that a higher $\frac{Heating\ volume}{Storage\ volume}$ >> 0.008 necessites to achieve these requirements, which in turn will improve the flow rate of the HTF in the thermosyphon loop.

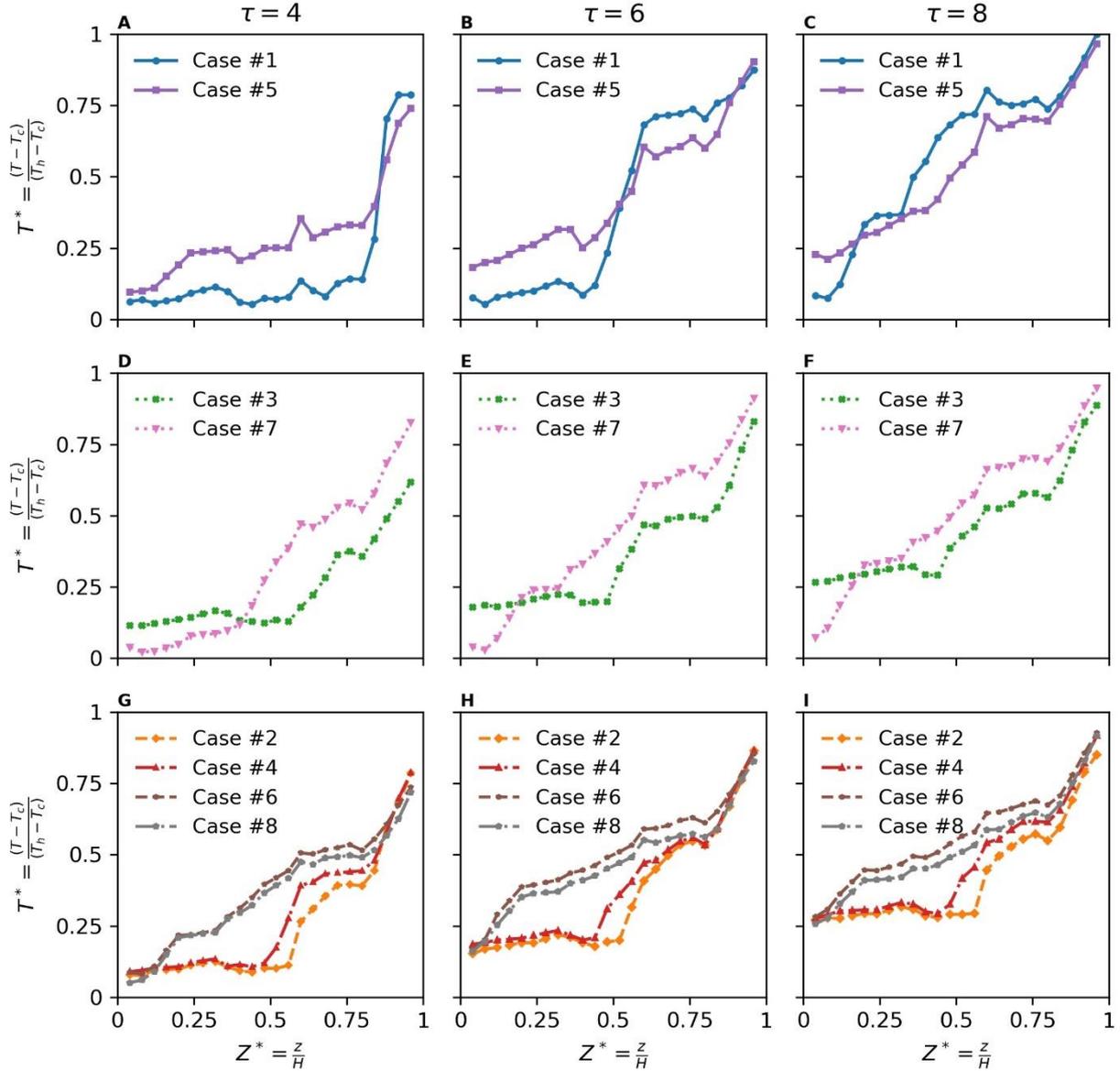

Figure 9: Varying thermal stratification in the storage tank during charging; Normalized temperature profiles $T^*$ plotted against height $Z^*$ at $\tau = 4, 6,$ and 8.

### 3.1.4. Charging efficiency

To analyze thermosyphon charging, the study cases are compared with respect to their charging efficiencies. As described in Mawire et al. [26], the charging efficiency is based on the First law of thermodynamics and calculated by

$$\eta(t) = \frac{T_{avg}(t) - T_{ini}}{T_{inlet}(t) - T_{ini}} \tag{6}$$

where $T_{avg}(t)$ is the time-dependent average temperature of the thermocline storage. $T_{inlet}(t)$ is the inlet temperature of the hot HTF (same as the jacketed heater outlet temperature as indicated in Figure 6), and $T_{ini}$ is the average temperature of the storage at the beginning of the charging, i.e., at $\tau = 0$. The time-varying efficiencies of the case studies are depicted in Figure 10.

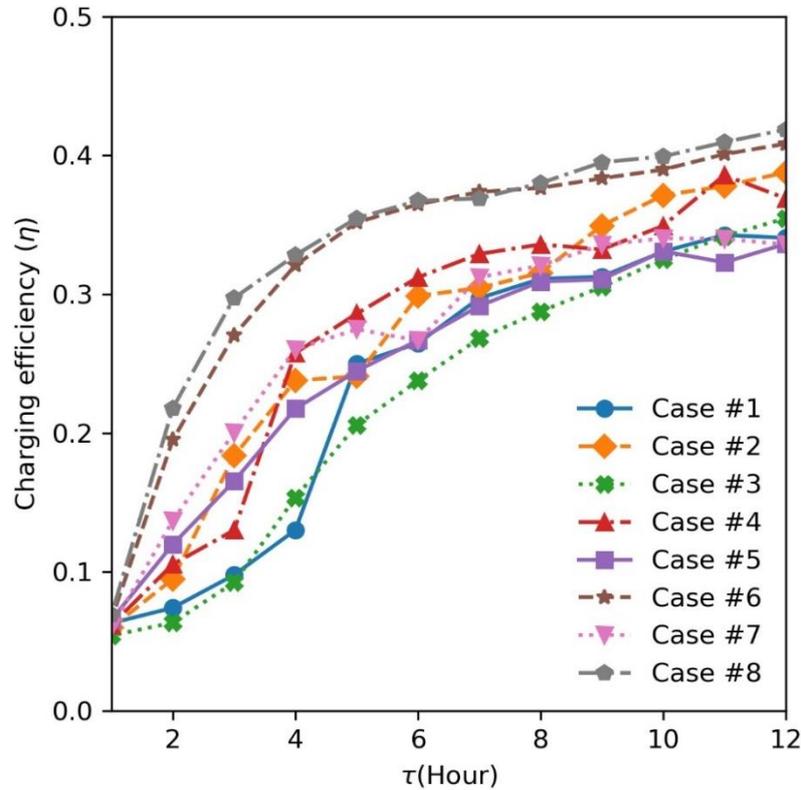

Figure 10: Charging efficiencies of the stratified storage for all study cases at one-hour intervals.

As illustrated in the figure, the charging efficiency ($\eta$) increased rapidly up to 5 hours and then slowed down. At the end of charging, i.e., at $\tau = 12$, the maximum $\eta$ is found to be ~0.4 for cases #6 and 8; ~0.35 for cases #2 and 4; and ~0.33 for the rest of the cases. In addition, the $\eta$ profiles are overlapped, except for cases #6 and 8. This suggests that, firstly, the charging efficiency increases for high-temperature cases. Secondly, the bottom-expansion design is considerably more effective than the top-expansion design.

### 3.2. Storage period
#### 3.2.1. Temperature profiles

The variation in temperature profiles during the storage period is illustrated in Figure 11 for the low-temperature cases (150 °C) and Figure 12 for relatively high-temperature cases (250 and 300 °C). The temperature readings and the height of the thermocouples are normalized according to equations 4 and 5, respectively. The temperature profiles are plotted at 6-hour intervals for 24 hours. A dwell time of 30 minutes is given after switching off the jacketed heater to account for any residual flow of HTF. Thus, the profiles corresponding to $\tau = 0$ merely indicate the final profiles at the end of charging or the beginning of the storage period. As seen in Figure 11, the stratified temperature profile is maintained for the entire storage period for the low temperature (150 °C) studies. Moreover, a subtle difference in the temperature profiles is observed between Case #1 and Case #5, and the degradation in the first 6 hours is very minimal. It suggests that for low-temperature charging, the resulting axial temperature of thermocline storage at the end of charging and their degradation during the storage are independent of the expansion-connection designs.

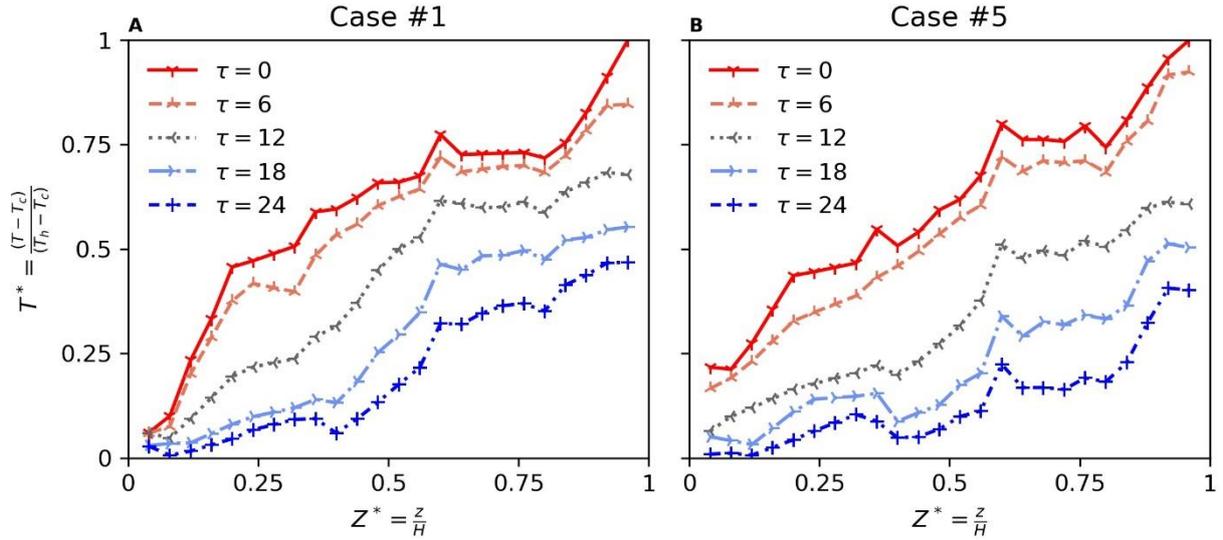

Figure 11: Temporal degradation of thermal stratification inside the storage (normalized temperature $T^*$ versus height $Z^*$) for 24 hours with 6-hour gap; A and B are low temperature (150 °C) continuous charging cases corresponding to the top and bottom expansion designs, respectively.

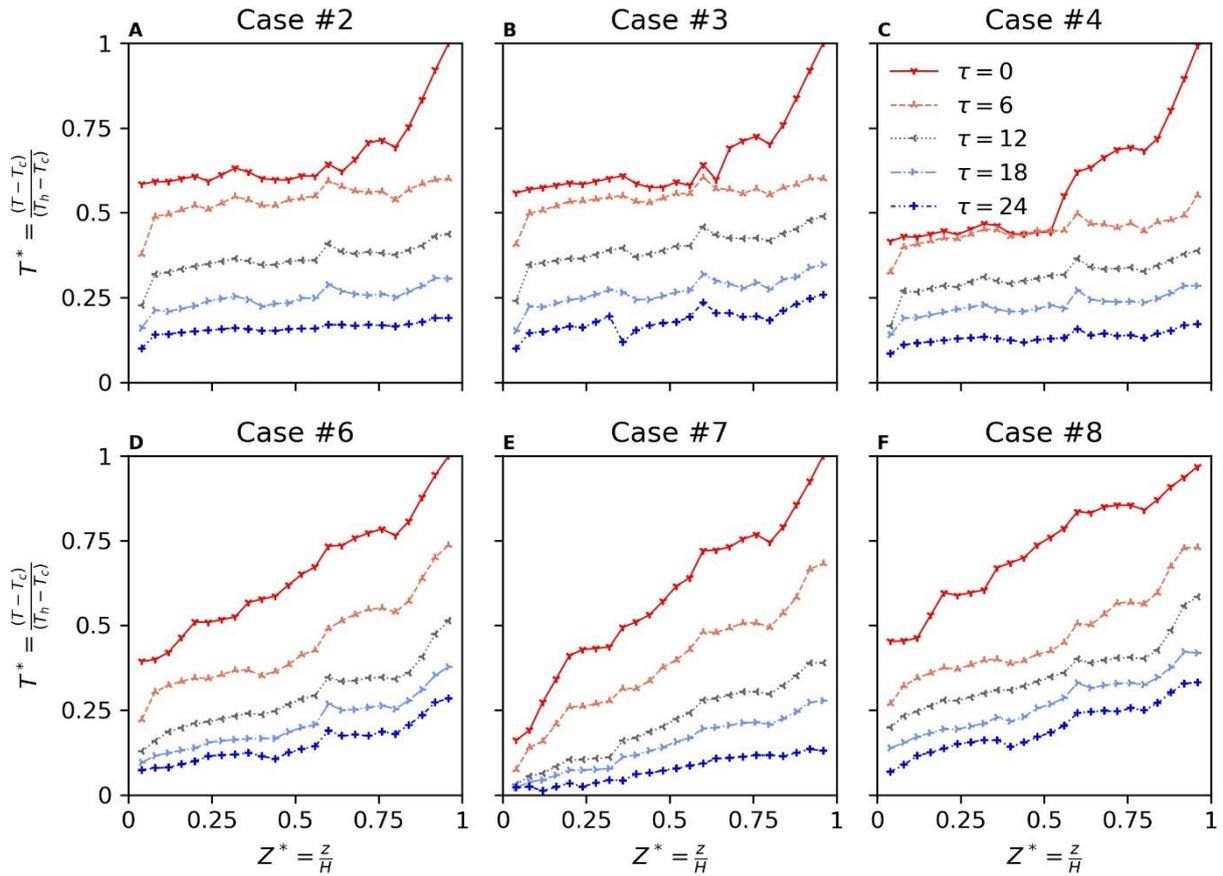

Figure 12: Temporal degradation of thermal stratification ($T^*$ versus $Z^*$ plotted for 24 hours with a 6-hour gap) for high-temperature study cases; A, B, and C correspond to top-expansion design, and D, E, and F correspond to bottom-expansion design; Cases #3 and 7 are studies for 250 °C charging, and the rest cases are for 300 °C charging.

However, the high temperature (250 and 300 °C) charging studies give contrasting results. As observed in Figure 12 (A, B, and C), the top-expansion design does not effectively maintain a clear stratification at the end of charging. Moreover, the bottom half of the storage is filled by thoroughly mixed HTF (i.e., $T^* \approx 0.5$), and any stratification present near the top of the storage tank is degraded during the first 6 hours of the storage period. For bottom-expansion design, clear thermal stratification is observed in Figure 12 (D, E, and F) at the beginning, $\tau = 0$. But, the degradation during first 6-hour is relatively larger. This suggests that the heat loss due to wall conduction as well as to the surrounding is significant at higher temperatures. Nevertheless, the thermal stratification is sustained till the end of the storage period.

### 3.2.2. Comparison of thermal stratification

As discussed earlier, thermal stratification is the basis of single-tank sensible heat storage, and thermosyphon charging is employed to reduce instrument and operational costs. Hence, it is essential to assess the resulting thermal stratifications at the end of the charging as well as their temporal degradation. This has been estimated using a quantitative index named Ideal Stratification Index (ISI) [7], which is defined by

$$ISI = \frac{\sum_{j=1}^{n} \theta_j^* \times f(\theta_j^*)}{n} \tag{7}$$

where, $\theta_j^*$ is the equivalent normalized temperature ($\theta_j^* = 0.5 + |T_j^* - 0.5|$) corresponding to $j^{th}$ thermocouple, $n$ is the total number of thermocouples and $f(\theta_j^*) = 4(\theta_j^* - 0.5)^2$ is the ideality factor. In particular, ISI analysis quantitatively yields how nearly any real stratification profile approaches the ideal stratification. The ISI values vary between 0 (for fully mixed HTF, i.e., $T^* = 0.5$) and 1 (for unmixed HTF, i.e., $T^* = 0$ or 1). For clarity, a schematic of ideal and real stratifications is shown in Figure 13 A, and the temporal degradation of ISI in the rest of Figures B, C, and D.

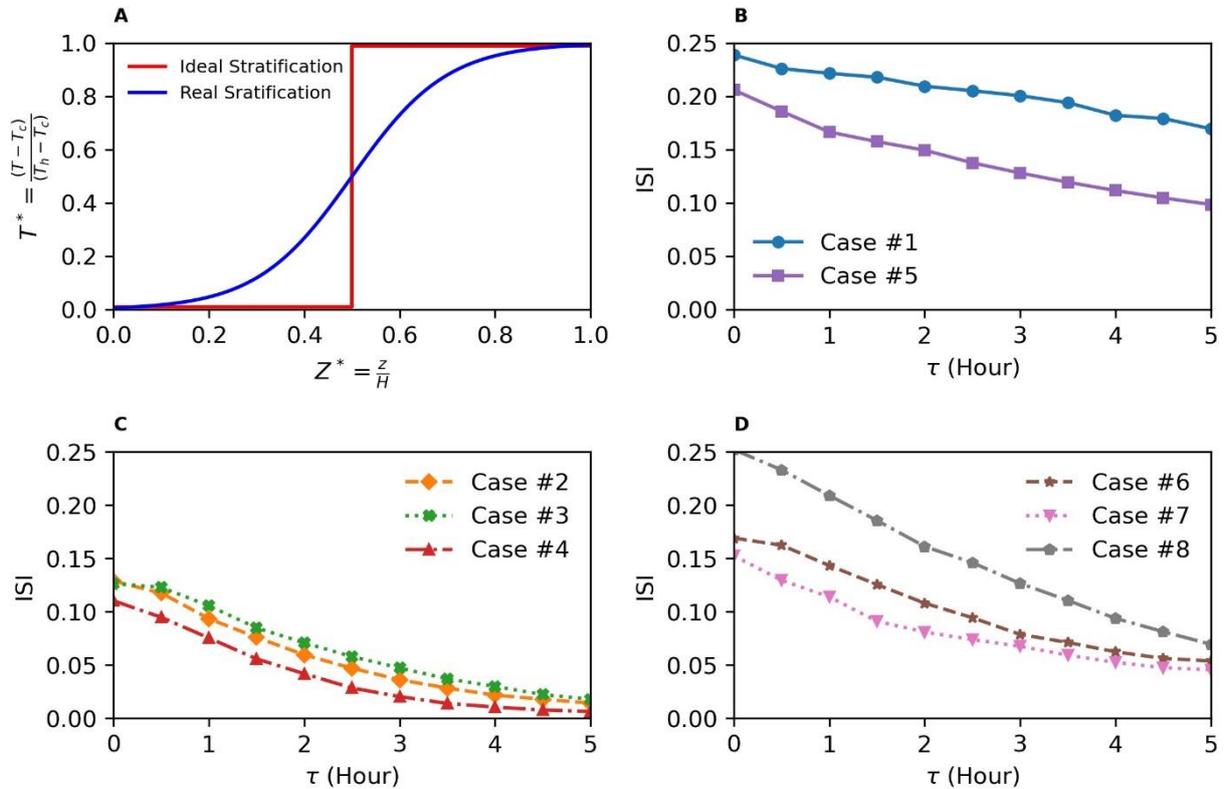

Figure 13: Comparison of thermal stratification with respect to ideal stratification; **A** represents a schematic of ideal and real stratified profiles at an instant; **B**, **C**, and **D** show the temporal degradation of the stratified profiles.

The comparative ISI graphs are plotted only for the first 5 hours because the quantitative index for the high-temperature study cases (Figure 13 C and D) achieved the lower limit at around $\tau = 5$. In addition, the resulting stratification in the thermocline storage appeared up to a maximum value of ~0.25 (as indicated in cases #1 and 8). However, the slope of the ISI curve for case #8 is steeper than that of case #1. This suggests that temporal degradation of the thermocline is faster in high-temperature studies. Moreover, a comparison of ISI curves between Figure 13 C and D reveals that the bottom-expansion design performs relatively better than the top-expansion design. The above discussion suggests firstly that the maximum ideal stratification in thermosyphon charging has a design limit (approx. 0.25 for the present loop) irrespective of thermal expansion design and continuous/pulsating charging. Secondly, there is an operational limit, which is the maximum layover period, preferable, between intermittent charging of the TES (5 hours for the present loop). For CSP plants, the thermosyphon-charging of the stratified TES may be interrupted due to bad weather, and the energy loss from the TES will be significant if the charging is delayed more than this limit.

### 3.2.3. Exergy degradation

Regarding thermodynamic analysis, exergy degradation is one of the most important parameters to evaluate the performance of any thermal energy storage. Exergy is calculated from the Second law of thermodynamics. The degradation of exergy indicates the loss of extractable energy in the storage system. Since spatial temperature variation is present in the stratified storage, the exergy of the whole TES is equal to the sum of the exergy of individual zones formed around the thermocouples. As reported in Advaith et al. [6], the exergy of the stratified tank is calculated by

$$\xi_t = \sum_{i=1}^{n} \rho_i V_i C_{p_i} [(T_{i,t} - T_0) + T_0 \ln\left(\frac{T_0}{T_{i,t}}\right)] \tag{8}$$

where $\rho_i$, $V_i$, $C_{p_i}$, and $T_{i,t}$ are the density, the volume, the specific heat capacity, and the temperature reading of the HTF corresponding to an i$^{th}$ zone. The temperature-dependent thermophysical properties ($\rho$ and $C_p$) are calculated from Equations (1) and (2). $T_0$ is the ambient temperature, and $n$ is the total number of zones. For this study, $T_0 = 25$ °C, and $n = 24$. For comparison, the exergy of the TES at 6-hour intervals is normalized with respect to the exergy at the beginning of the storage (at $\tau = 0$), i.e., $\xi^* = \frac{\xi_t}{\xi_0}$ and illustrated in Figure 14.

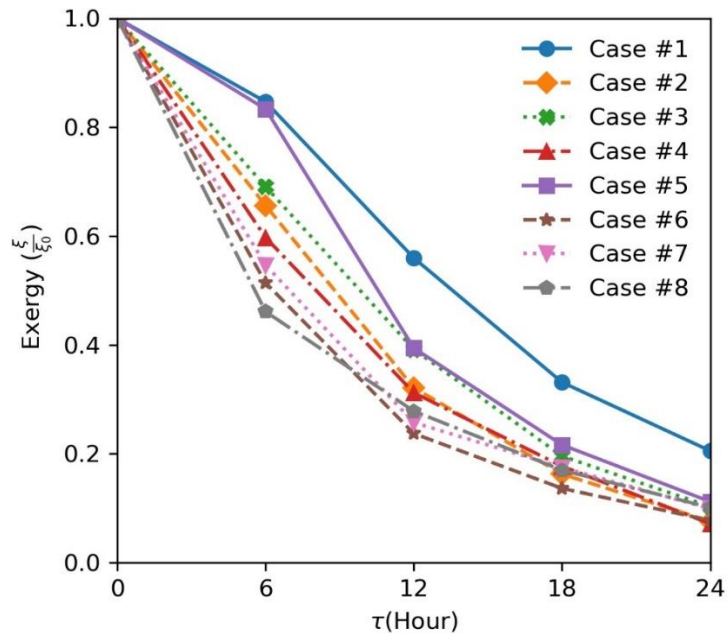

Figure 14: Exergy degradation of the stratified storage with respect to time during the storage period.

As indicated from the slope of the curves in Figure 14, the exergy decays faster at the beginning of the storage period, which then slows down with time. Moreover, the loss in exergy for low-temperature studies (cases #1 and 5) is relatively lower than in high-temperature studies (cases #2, 3, 4, 6, 7, and 8). In general, the exergy of a system degrades by both energy loss and internal entropy generation. For stratified TES, heat loss to the ambient corresponds to the energy loss, and internal mixing of hot and cold HTF (caused by axial diffusion and wall conduction) corresponds to the internal entropy generation. Since both these factors are significant for high-temperature storage, a higher loss in exergy ($\xi^*$) is observed compared to low-temperature storage.

In addition, the exergy losses for cases #2, 3, and 4 are less compared to cases #6, 7, and 8 for $\tau = 6$. Since the initial temperature profiles (at $\tau = 0$) were relatively less stratified for study cases #2, 3, and 4 than that for cases #6, 7, and 8 (see Figure 12), it suggests that in stratified storage, the rate of exergy degradation increases with the degree of stratification owing to internal entropy generation. Lastly, a significant exergy loss (~35 to 50%) is observed in the first 6 hours of storage for the high-temperature studies. It indicates that the maximum layover period for this stratified storage should be less than 6 hours, which corroborates with the results obtained from the ideal stratification index ($\leq 5$ hours).

As part of this effort, an experiment has been performed for the thermosyphon charging of a low-melting eutectic salt mixture known as the HITEC salt. Although the investigation is beyond the scope of this manuscript, the authors report associated challenges and safety issues faced in conducting a molten-salt experiment in the Appendix section.

## 4. Conclusion

An experimental investigation was carried out on a single-medium stratified thermal energy storage to assess thermosyphon charging for concentrated solar power applications. The stratified storage was cylindrical in shape with an aspect ratio $\frac{Height}{Diameter} \approx \frac{4}{1}$ and a storage volume of 370 liters. The ratio of the heating volume to storage volume for thermosyphon charging was 0.008, and thermal expansion of the HTF was accommodated in an expansion tank by two different designs (top and bottom expansion connections). Finally, continuous and pulsatile charging was carried out for 12 hours, followed by a storage period of 24 hours.

The results obtained from these experiments show that the bottom-expansion design yields a relatively larger temperature (~25 °C) for the HTF at the jacketed heater outlet than that of the top-expansion design for both continuous and pulsatile charging. Also, the Atwood number difference can be used as a parameter for decision-making on the design and operation of the thermosyphon charging loop. Furthermore, this investigation reveals that there is a design constraint to the ultimate stratification of HTF using thermosyphon charging as the resulting $A_t(TS)$ for the case studies are around ±0.045 irrespective of the type of expansion setup and operational variations.

Additionally, this study indicates that the HTF experiences multi-pass charging despite variations in operational settings for the low $\frac{Heating\ volume}{Storage\ volume}$ of 0.008. Therefore, increasing heating volume should be considered for designing a thermosyphon-charging loop for concentrated solar power applications. On top of that, this study suggests that there is a design limit to how much the stratified profiles are closer to ideal stratification for a thermosyphon loop (~ 0.25 for the present loop). Most of all, the results demonstrate that the charging efficiencies corresponding to the bottom-expansion design are considerably higher compared to the top-expansion design for high-temperature case studies. Finally, the storage period results indicate that the exergy decreases faster if the HTF is relatively more stratified in the TES, and the maximum layover period for the stratified storage can be determined from the degradation of the exergy and ideal stratification index.

To conclude, this study provides valuable insights into thermosyphon charging of single-medium stratified storage, particularly important from an operational perspective for high-temperature applications. Moreover, this study has a limitation concerning the heat loss to supporting structures which should be minimized in concentrated solar power plants by providing a concrete foundation, sand layers, ceramic or firebrick insulation, and so on. Lastly, future studies should be extended to higher temperature storage using molten-salt-based heat transfer fluids.

## 5. CRediT authorship contribution statement

**Dipti Ranjan Parida**: Conceptualization, Methodology, Experiments, Formal analysis, Writing - Original Draft

**Saptarshi Basu**: Conceptualization, Funding acquisition, Supervision, Writing - review & editing

**Dhanush A P**: Experiments, Data Curation

## 6. Declaration of Competing Interest

The authors declare that they have no known competing financial interests or personal relationships that could have appeared to influence the work reported in this paper.

## 7. Acknowledgments

This work is supported by the Ministry of Human Resource Development, and the Ministry of New and Renewable Energy, Government of India, under the IMPRINT initiative (Project No. 4424).

The authors thank Mr. Nikhil Dani (former project associate) and Murhopye Scientific Company Pvt. Ltd. for their assistance in setting up the experimental loop; and Dr. Prasenjit Kabi for his valuable comments that greatly improved the manuscript.